# Non-Centrosymmetric γ-Phase GaS Nanobelts for On-Chip Nonlinear Photonic Applications

*Yukihiro Endo[1*], Masato Takiguchi[1, 2], Yoshiaki Sekine[1], Hisashi Sumikura[1, 2], and Yoshitaka Taniyasu[1]*

[1]Basic Research Laboratories, NTT, Inc., 3-1 Morinosato Wakamiya, Atsugi, Kanagawa 243–0198, Japan

[2]NTT Nanophotonics Center, NTT, Inc., 3-1 Morinosato Wakamiya, Atsugi, Kanagawa 243–0198, Japan

E-mail: yukihiro.endo@ntt.com

**Abstract**

Second-order nonlinear optical processes in van der Waals (vdW) semiconductors offer a compelling route toward compact, integrable photon-conversion platforms. Among III–VI vdW semiconductors, GaS is particularly attractive owing to its wide bandgap suppressing two-photon absorption under near-infrared laser excitation. However, bulk GaS typically crystallizes in the centrosymmetric β phase, which eliminates second-order nonlinearity and severely limits its application in nonlinear photonics. Here, we demonstrate that GaS nanobelts synthesized via self-catalyzed vapor–liquid–solid growth predominantly crystallize in non-centrosymmetric γ-phase stacking. This behavior originates from edge-selective growth kinetics at the Ga catalyst interface, which stabilizes the γ phase and enables deterministic in-plane dipole moment alignment. The GaS nanobelts exhibit strong second-harmonic generation (SHG) with intensities comparable to those of GaSe, a widely used nonlinear optical material. Furthermore, we integrate the nanobelts onto SiN waveguides and demonstrate efficient on-chip SHG and sum-frequency generation. These results establish γ-GaS nanobelts as a transferable one-dimensional nonlinear materials well suited for on-chip photonic integration and indicate their strong potential for nonlinear optical applications.

## 1. Introduction

Nonlinear optical responses in van der Waals (vdW) materials have recently attracted significant attention owing to their large nonlinear susceptibilities,[1] quasi-phase-matching by vdW stacking,[2, 3] and excellent compatibility with heterogeneous photonic integration.[4–9] In particular, III–VI vdW semiconductors, such as GaSe and InSe, exhibit remarkably strong second-order nonlinearities, enabling efficient second-harmonic generation (SHG),[10–13] sum- and difference-frequency generation (SFG and DFG),[9, 14–16] optical parametric amplification (OPA),[17] and spontaneous parametric down-conversion (SPDC).[18] Among them, GaSe has already been commercially utilized as a nonlinear optical crystal for SHG of $CO_2$ lasers, broadband DFG-based MIR sources, and terahertz wave generation.

The large nonlinear response of these materials originates from their non-centrosymmetric atomic configurations. From a crystallographic perspective, each unit layer consists of nearest-neighbor group III and VI atomic pairs aligned along the armchair direction, resulting in efficient coupling with the electric field of incident light. Furthermore, their excellent exfoliability enables facile integration into various optical platforms, including photonic crystals,[5, 8] photonic circuits,[7] and optical fibers,[9, 12] thereby opening promising opportunities for next-generation nonlinear photonic devices.

Among III–VI vdW semiconductors, GaS possesses the largest bandgap energy (~2.5 eV), extending the short-wavelength cutoff compared with GaSe (~2.0 eV) and InSe (~1.3 eV). The larger bandgap energy is advantageous for suppressing two-photon absorption (TPA) during second-order nonlinear optical processes under near-infrared (NIR) laser pumping. Moreover, theoretical calculations predict that GaS exhibits a large second-order nonlinear susceptibility $\chi^2$ (62.4 pm/V), comparable to that of GaSe (64.6 pm/V).[19] Despite these attractive properties, GaS has rarely been used as a nonlinear optical material due to its stacking structure.

Bulk GaS predominantly crystallizes in the β-phase polytype, which possesses a centrosymmetric stacking configuration and therefore lacks second-order nonlinear optical response. Previous efforts to circumvent this limitation have mainly focused on alloying strategies, particularly $GaS_xSe_{1-x}$, where non-centrosymmetric stacking phases can be stabilized for limited sulfur concentrations ($x \leqq 0.4$).[20–24] Such alloying approaches were originally motivated by the need to suppress TPA under NIR excitation, especially for widely used 1064-nm pump lasers, by enlarging the bandgap relative to GaSe. Although $GaS_xSe_{1-x}$ alloys exhibit reduced TPA and improved laser-induced damage thresholds compared with GaSe,[21] their bandgaps remain smaller than the two-photon energy (2.3 eV) under 1064-nm excitation,[25] preventing efficient suppression of TPA.[26] More critically, the introduction of sulfur at levels as low as ~19% induces the coexistence of the β phase in the alloys,[27] leading to randomization of the in-plane dipole moment orientations and thereby drastically degrading nonlinear optical performance. These limitations motivate alternative approaches that directly stabilize the non-centrosymmetric stacking order in GaS itself.

In this context, recent work by Y. Song *et al*. has provided an important perspective by demonstrating the observation of SHG in GaS.[28] By exfoliating bulk GaS crystals, they obtained a large number of GaS flakes, and approximately 14% of them exhibited SHG signals. Through systematic SHG measurements, they revealed that although the β phase is the dominant polytype, a small fraction of the γ phase coexists within the GaS crystals. This finding indicates that the thermal stabilities of the β and γ phases in GaS are very close. Consequently, for practical nonlinear optical applications, it is highly desirable to stabilize the non-centrosymmetric γ phase of GaS as a dominant phase during growth while simultaneously achieving a macroscopically uniform in-plane dipole moment orientation.

In parallel, nanobelt (or nanoribbon) structures have emerged as especially promising building blocks for integrated photonics.[29–32] Their elongated geometry and uniform thickness make them inherently compatible with waveguide-based architectures. More specifically, integration of nanobelts into waveguides enables efficient optical mode overlap and the incorporation of additional functionalities into photonic circuits.[31] Compared with conventional cylindrical shaped nanowires, flat belt-shaped nanobelts offer superior integration compatibility with planar photonic elements owing to their wide and flat surfaces, which facilitate intimate and low-loss coupling. In addition, nanobelts based on vdW materials can be integrated using exfoliation and transfer processes similar to those used for conventional two-dimensional vdW materials, thereby expanding their compatibility with a wide range of photonic platforms.

Recently, we successfully synthesized GaS nanobelts via self-catalyzed vapor–liquid–solid (VLS) growth and found that they predominantly crystallize in the γ-phase polytype.[32] More recently, J. Jing *et al.* also reported the predominance of the γ phase in GaS nanobelts.[33] These findings suggest that GaS nanobelts are promising materials for nonlinear nanophotonic applications. Despite these advantages, however, the mechanism underlying γ-phase formation remains unclear, and the on-chip optical integration of GaS nanobelts has not yet been explored.

In this work, we demonstrate the growth of GaS nanobelts crystallized in the non-centrosymmetric γ-phase stacking structure using a self-catalyzed VLS growth technique. We show that γ-phase formation can be explained by edge-selective growth kinetics at the Ga catalyst interface. Consequently, all GaS nanobelts exhibit second-harmonic generation, in contrast to conventional bulk GaS crystals. We further integrate GaS nanobelts onto SiN waveguides and demonstrate on-chip nonlinear optical functionality.

## 2. Results and Discussion

### 2.1. Growth of γ-GaS Nanobelts

GaS nanobelts were synthesized via self-catalyzed VLS growth using metalorganic chemical vapor deposition (MOCVD) (see experimental section for details). **Figure 1**(a) shows an optical microscope image of GaS nanobelts grown on a sapphire substrate, and Figure 1b presents a close-up image of a GaS nanobelt. One end of the nanobelt is terminated by a Ga particle, consistent with the self-catalyzed VLS growth mechanism.[32] As shown in Figure 1(a), three distinct growth orientations are observed along the $[\mathbf{11\bar{2}0}]$, $[\mathbf{\bar{1}2\bar{1}0}]$, and $[\mathbf{2\bar{1}\bar{1}0}]$ directions of the sapphire crystal. This clearly indicates the epitaxial growth of GaS nanobelts, with their growth directions determined by the epitaxial relationship between the sapphire c-plane and the GaS nanobelts, as illustrated in Figure 1(c). To the best of our knowledge, this is the first demonstration of in-plane epitaxial growth of GaS on c-plane sapphire, opening up opportunities for a wide range of applications enabled by highly aligned GaS nanostructures.

To evaluate the morphology and surface flatness of the GaS nanobelt shown in Figure 1(b), atomic force microscope (AFM) measurements were performed, as presented in Figure 1(d). The nanobelt had a height of 230 nm and a width of ~2.5 µm. The inset shows a topographic image of the nanobelt surface. The absence of atomic steps and the low root-mean-square roughness (230 pm) indicate an atomically flat surface. Figure 1(e) shows line profiles crossing the nanobelt along the dashed lines in Figure 1(b). The identical heights observed in these profiles demonstrate that the GaS nanobelt possesses a flat surface over a wide area exceeding 100 µm in length. Such thickness uniformity is a significant advantage for nanophotonics applications.

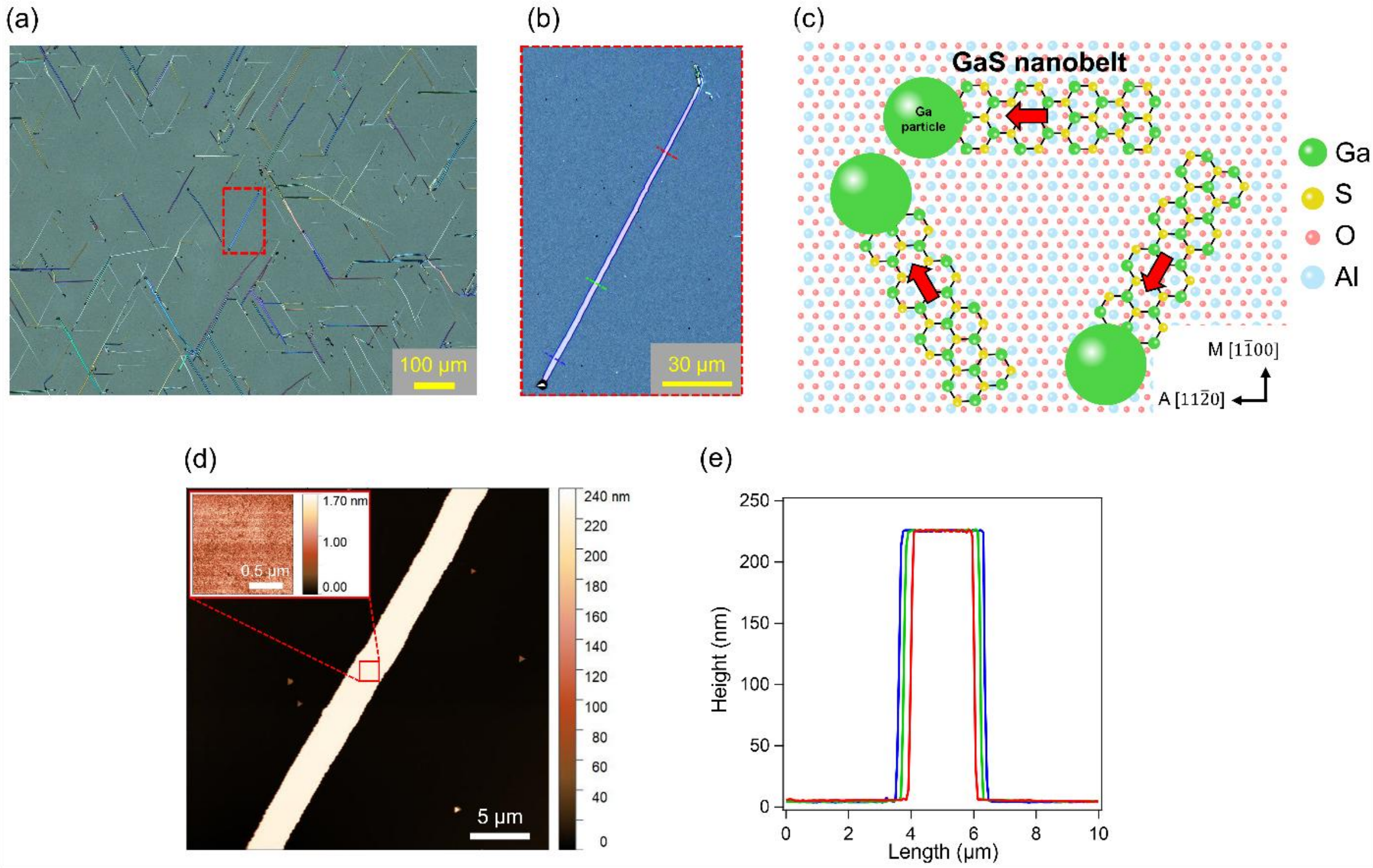

**Figure 1.** Growth and morphology of GaS nanobelts. (a) Optical microscope image of GaS nanobelts grown epitaxially on sapphire substrate, showing three preferential growth orientations. (b) Magnified image of GaS nanobelt [red dashed rectangle in (a)]. (c) Schematic illustration of epitaxial relationship between sapphire c-plane and GaS nanobelt. (d) AFM image of GaS nanobelt shown in (b). Inset shows magnified topographic image of nanobelt surface. (e) AFM line profiles along dashed lines in (b) indicate thickness uniformity over more than 100 μm. Colors of line profiles correspond to those of dashed lines in (b).

Next, we investigate the atomic arrangement, particularly the stacking order, and compare the GaS nanobelts with conventional GaS flakes exfoliated from bulk crystals (purchased from 2D semiconductors). The exfoliated GaS flakes were transferred onto sapphire substrate. **Figure 2**(a) illustrates the cross-sectional atomic arrangements of γ- and β-phase polytypes. These polytypes share the same unit layer but differ in their stacking sequences. The γ phase consists of three-unit layers, whereas the β phase consists of two-unit layers. In the γ phase, all unit layers stack with the same in-plane orientation, resulting in a non-centrosymmetric crystal structure with point group $C_{3v}$. In contrast, in the β phase with point group $D_{6h}$, the upper unit layer is rotated by 60° about the c-axis relative to the lower unit layer, giving rise to centrosymmetry.

Figures 2(b) and 2(c) show high-angle annular dark-field scanning transmission electron microscopy (HAADF-STEM) images of a GaS nanobelt and an exfoliated GaS flake, respectively. From these atomic-resolution images, each polytype was clearly identified. The GaS nanobelt exhibits the γ phase, and the exfoliated GaS flake is in the β phase. As shown in Figure 2(d), the diffraction pattern of the GaS nanobelt displays a bright 012 spot attributed to the γ phase, indicating that the γ phase is the dominant polytype in the GaS nanobelt. The diffraction pattern of the exfoliated GaS flake (Figure 2(e)) exhibits a 011 spot attributed to the β phase, confirming β as the dominant polytype.

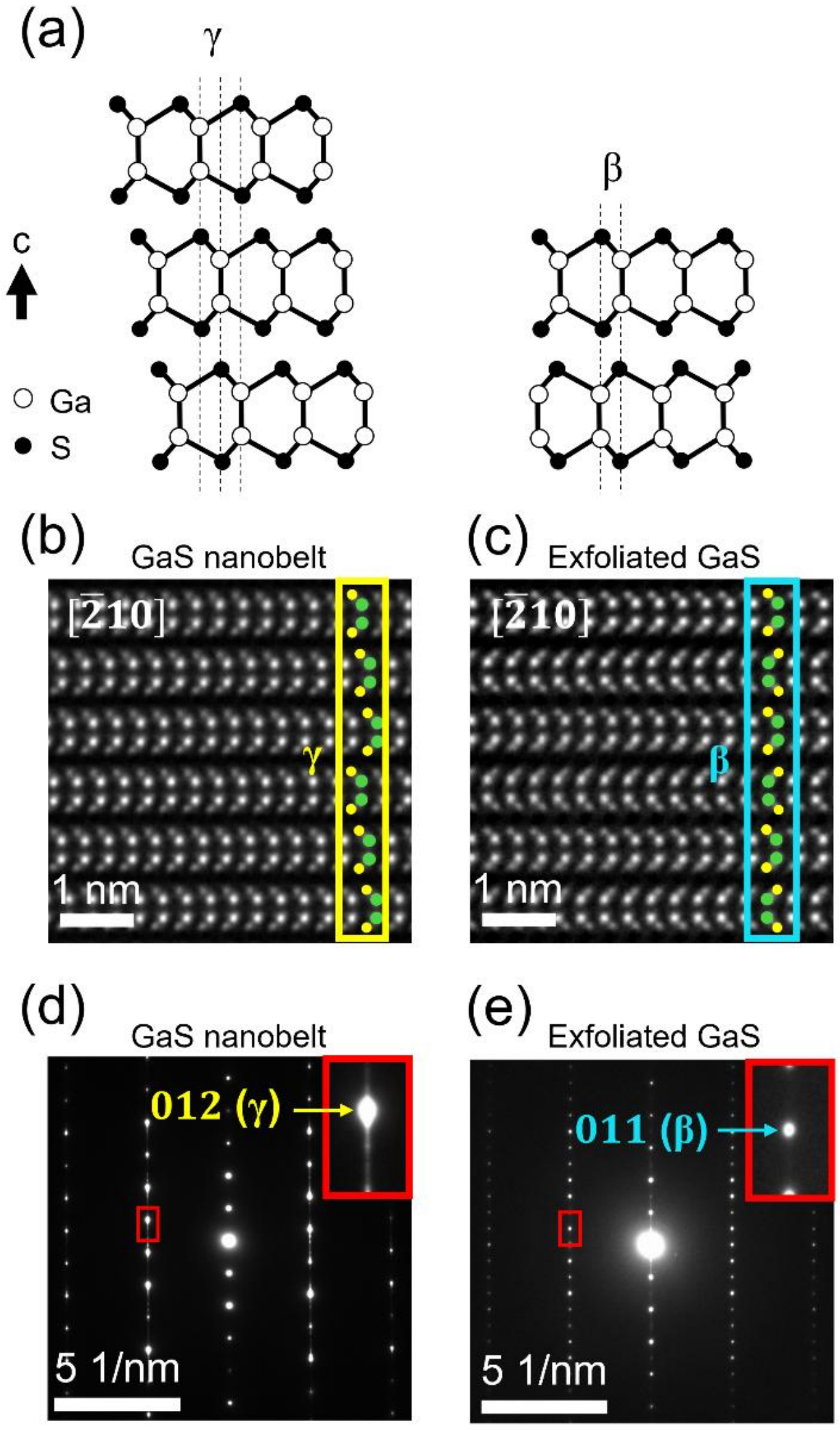

**Figure 2**. Stacking structure of GaS nanobelt compared with exfoliated GaS flake. (a) Schematic cross-sectional atomic arrangements of γ- and β-phase polytypes of III–VI atomic layered semiconductors. (b, c) HAADF-STEM images and (d, e) electron diffraction patterns of GaS nanobelt and exfoliated GaS flake, respectively. Insets of (d, e) show magnified diffraction spots surrounded by red rectangles in each diffraction pattern.

### 2.2. Deterministic In-Plane Dipole Moment Alignment

Next, we examine the stacking distribution across the GaS nanobelt thickness and elucidate the mechanism responsible for the preferential formation of the γ phase instead of the conventional β phase. **Figure 3**(a) shows a cross-sectional transmission electron microscopy (TEM) image of a representative GaS nanobelt. To evaluate the local stacking order, high-resolution HAADF-STEM images were acquired from the near-surface region (light-green dashed rectangle), the middle region (blue dashed rectangle), and the near-substrate region (pink dashed rectangle), as shown in Figure 3(b–d), respectively. The γ phase is dominant throughout the nanobelt, while a small fraction of β-phase stacking faults is observed in the middle and near-substrate regions (indicated by light-blue arrows in Figure 3(c,d)). Quantitative analysis revealed that the γ-phase fraction increased from 88% near the substrate to 100% near the surface. Importantly, despite the presence of these β-phase insertions near the substrate, the in-plane orientation of the GaS unit layers remained highly oriented to a single direction, as clearly seen in the magnified HAADF-STEM images in Figure 3(e–g).

This unusual robustness of the in-plane orientation can be understood in terms of the self-catalyzed VLS growth mechanism. Figure 3(h) illustrates the atomic configuration of the GaS nanobelt together with the Ga catalyst particle, corresponding to the experimentally observed geometry. As schematically shown in Figure 3(i), GaS exhibits three types of edges: armchair edge, 1-bond zigzag edge, and 2-bond zigzag edge. The armchair edge consists of Ga and S atoms in a 1:1 ratio, whereas zigzag edges are terminated exclusively by either Ga or S atoms. During self-catalyzed VLS growth, the Ga catalyst particle terminates the growth front, suppressing the formation of armchair edges and favoring zigzag-type interfaces.[32] Among the Ga-terminated zigzag edges, the 2-bond zigzag edge provides a higher density of dangling bonds and therefore a stronger affinity for incoming S sources. Because S supply limits the growth rate, this energetically favorable 2-bond zigzag edge preferentially defines the growth direction. As a result, the GaS nanobelt grows predominantly along the outward growth direction normal to this edge, as indicated by the red arrow in Figure 3(i). Figure 3(j) illustrates two unit layers of the γ phase and the β phase. We consider the configuration corresponding to the HAADF-STEM images, where the left edges (surrounded by red rectangles) represent the growth fronts (or Ga catalyst interfaces), and the first layer (1L) has a 2-bond zigzag edge at its growth front. In the γ phase, the growth front consistently maintains the 2-bond zigzag edge during further stacking. In contrast, in the β phase, the second layer (2L) exhibits a 1-bond zigzag edge at the growth front, and 1-bond and 2-bond zigzag edges appear alternately during subsequent stacking. Thus, the edge-selective growth kinetics at the growth front leads to a highly uniform in-plane orientation together with stabilization of the non-centrosymmetric γ phase, which results in deterministic in-plane dipole moment alignment across the entire nanobelt (the dipole moments arising from the in-plane nearest-neighbor Ga–S atomic pairs).

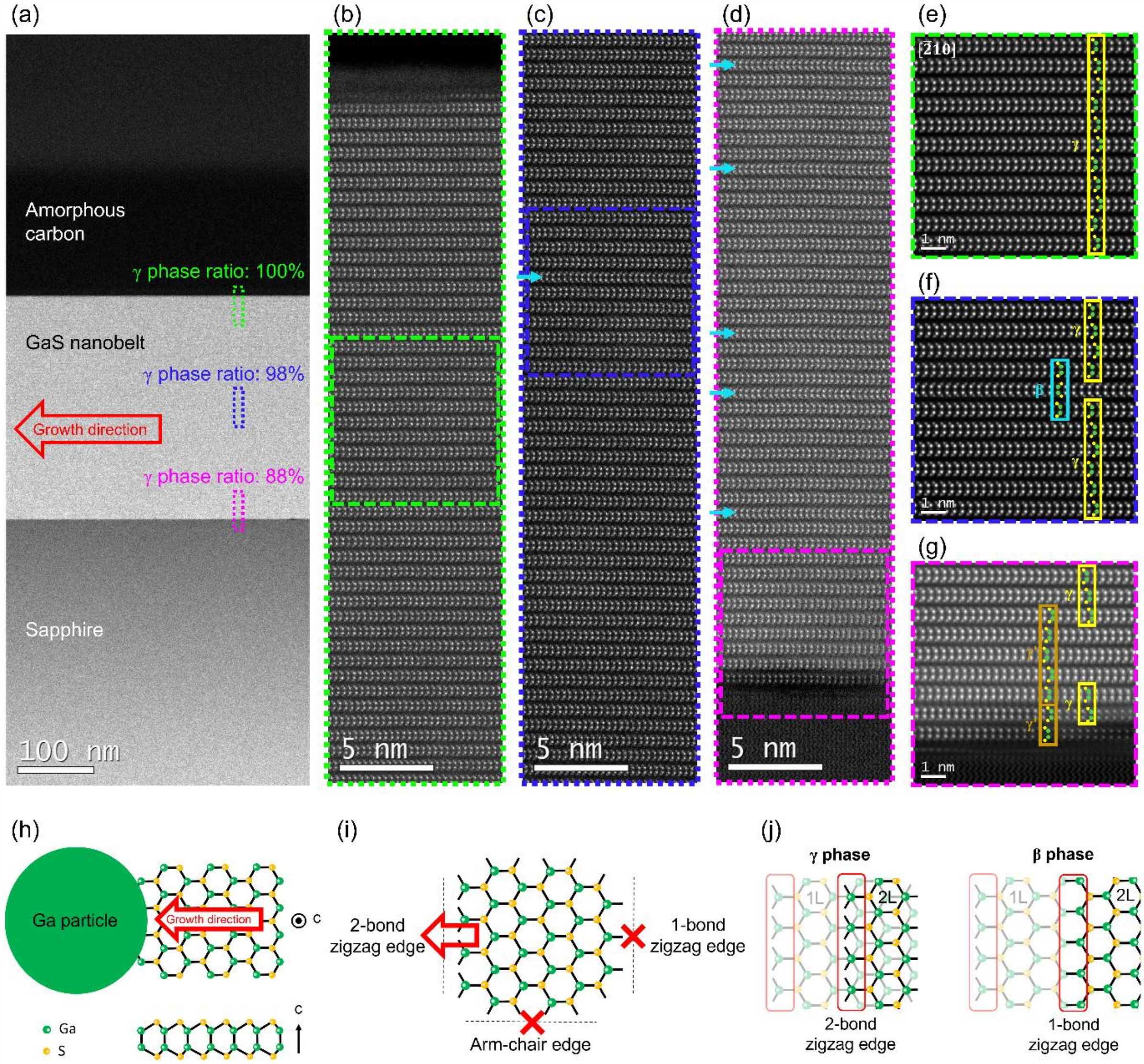

**Figure 3**. Stacking structure distribution and growth mechanism of GaS nanobelt. (a) Cross-sectional TEM image of GaS nanobelt. (b–d) HAADF-STEM images on near-surface region (light-green dashed rectangle in a), middle region (blue dashed rectangle in a), and near-substrate region (pink dashed rectangle in a), respectively. (e–g) High magnification HAADF-STEM images corresponding to dashed rectangles in (b–d), respectively. For convenience, we denote γ phase with reversed stacking order as γ' phase in (g) (note that in-plane orientations of γ and γ' phases are identical). Schematic illustrations of (h) atomic arrangement of GaS nanobelt together with Ga catalyst particle, and (i) three types of edges in GaS: armchair, 1-bond zigzag, and 2-bond zigzag edges. Red arrow indicates predominant growth direction of GaS nanobelt. (j) Schematic illustrations of two unit layers of γ phase (left) and β phase

(right). Growth fronts (or Ga catalyst interfaces) are surrounded by red rectangles. In γ phase, second layer (2L) maintains 2-bond zigzag edge at growth front. In contrast, in β phase, 1-bond zigzag edge appears at growth front.

### 2.3. Robust Second-Order Nonlinearity

To verify the second-order nonlinear optical response of the GaS nanobelts, SHG measurements were performed using a linearly polarized femtosecond laser with a fundamental wavelength of 1500 nm in a normal-incidence reflection geometry (see experimental section and **Figure** S2 for details). The generated SHG signals were detected around 750 nm. For comparison, exfoliated GaSe and GaS flakes (purchased from 2D Semiconductors) with similar thicknesses of approximately 140 nm were measured under identical conditions. The excitation polarization angle was adjusted to maximize the SHG intensity for each sample.

As shown in **Figure 4**(a), the GaS nanobelt exhibited a pronounced SHG signal with an intensity of 57.5 ± 0.4 cps, comparable to that of the exfoliated GaSe flake (48.4 ± 0.3 cps), which is widely used in nonlinear optics. In contrast, no detectable SHG signal was observed from the exfoliated GaS flake, consistent with its centrosymmetric β-phase crystal structure. Notably, SHG measurements performed on 28 randomly selected GaS nanobelts revealed clear SHG emission from all samples (see Supporting Information), demonstrating excellent sample-to-sample reproducibility. This behavior stands in stark contrast to exfoliated GaS flakes, for which significant variability in the SHG response has been reported.[28]

To further characterize the SHG properties of the GaS nanobelt, we measured the dependences of the SHG intensity on the excitation power and the crystal orientation. As shown in Figure 4(b), the SHG intensity exhibited an approximately quadratic dependence on the excitation power, following $I_{SHG} \propto P^{\alpha}$ with $\alpha = 2.1 \pm 0.3$, which is characteristic of a second-order nonlinear optical process. Figure 4(c) shows the SHG intensity as a function of the crystal rotation angle, measured with the excitation polarization fixed parallel to the collected SHG polarization component. A clear sixfold rotational pattern is observed, reflecting the threefold rotational symmetry of the non-centrosymmetric γ phase. These results unambiguously confirm that the GaS nanobelts possessed strong and robust second-order nonlinear optical activity arising from their γ-phase stacking. Although a minor fraction of centrosymmetric β-phase stacking is present, its influence on the overall nonlinear optical response is negligible because the macroscopic in-plane orientation is well preserved, as discussed above. Consequently, the GaS nanobelts maintained a strong second-order nonlinear optical response and effectively behaved as nearly single-crystalline, non-centrosymmetric structures, making them highly promising for nonlinear optical applications.

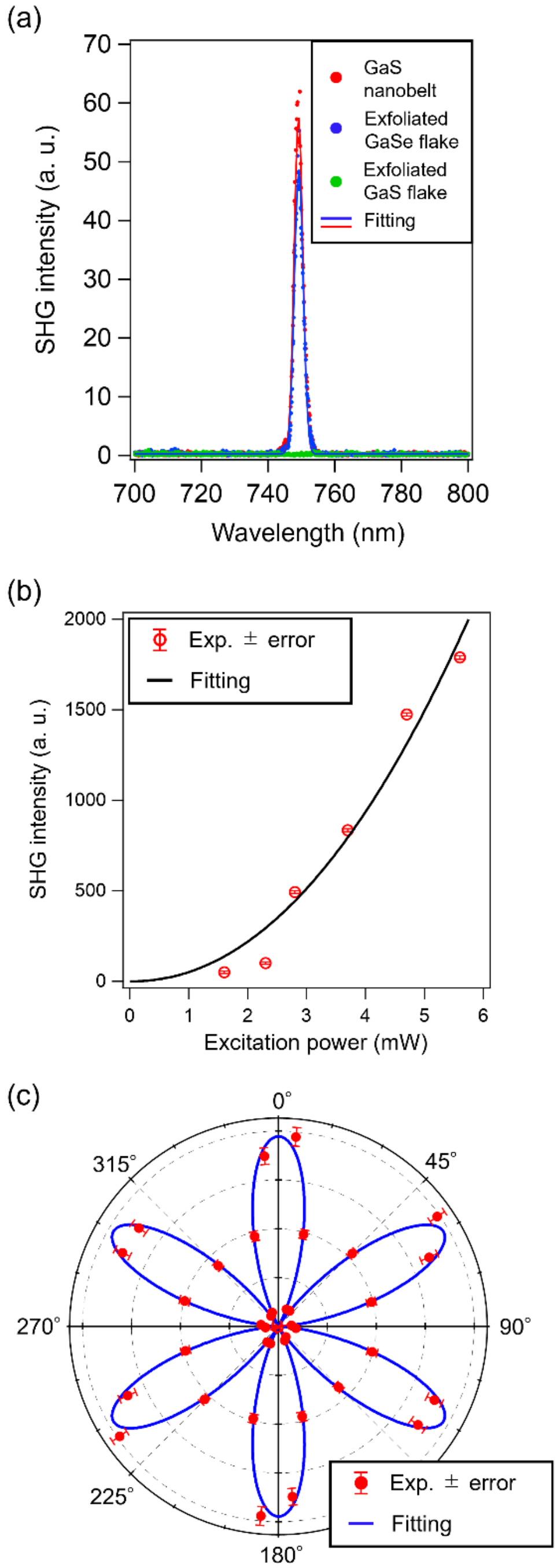

**Figure 4.** SHG from GaS nanobelt. (a) SHG intensities measured from GaS nanobelt, exfoliated GaSe flake, and exfoliated GaS flake under identical excitation conditions (wavelength of 1500 nm), showing SHG from non-centrosymmetric samples (GaS nanobelt and GaSe flake) only. (b) SHG intensities as function of excitation power, showing approximately quadratic dependence characteristic of second-order nonlinear optical process. (c) SHG intensities as function of crystal rotation angle with excitation polarization fixed parallel to collected SHG polarization component. At 0°, excitation polarization is parallel to long axis of GaS nanobelt. Observed sixfold rotational pattern consistent with threefold rotational symmetry of γ phase of GaS nanobelt.

### 2.4. Waveguide-Integrated Nonlinear Conversion

We integrated the GaS nanobelt onto a SiN waveguide to form an on-chip nonlinear optical waveguide, as shown in **Figure 5**(a). The GaS nanobelt was transferred onto the SiN waveguide using a polymer-assisted transfer technique (see experimental section for details). For low-power and low-cost practical applications, continuous-wave (CW) operation is highly desirable. To evaluate the feasibility of CW-driven nonlinear optical processes in the integrated platform, the SHG characteristics of the GaS nanobelt-integrated waveguide were first investigated using a CW excitation source. Figure 5(b,c) presents an optical microscope image and the corresponding SHG emission image acquired from the same observation area, respectively. To obtain the SHG emission image shown in Figure 5(c), the fundamental CW light (wavelength ~1500 nm) was coupled into the SiN waveguide from the chip edge, as illustrated in Figure 5(a), and the generated SHG signal (wavelength ~750 nm) was collected from the out-of-plane direction using a high-resolution CCD camera with a detectable wavelength range from visible to 1100 nm. A clear SHG emission is observed exclusively from the region where the GaS nanobelt is integrated on the SiN waveguide, indicating efficient optical coupling between the GaS nanobelt and the guided modes of the SiN waveguide.

To further confirm the optical coupling between the GaS nanobelt and the SiN waveguide, numerical simulations were performed using COMSOL. In the simulations, the thicknesses of the GaS nanobelt and SiN waveguide were set to 100 nm and 350 nm, respectively, while their widths were 2 μm and 3 μm. These values are based on the AFM line profile (Figure (5d)) taken along the dashed line in Figure 5(b). As shown in Figure 5(e), the simulated power flow distribution clearly indicates that the light was well confined within the waveguide, and the optical confinement factor in the GaS nanobelt was evaluated to be 11%. Although the GaS nanobelt was not positioned at the center of the SiN waveguide, the propagating light followed the position of the GaS nanobelt. This behavior arises because GaS ($n_o$ = 2.6, $n_e$ = 2.4 at 1550 nm) has a higher refractive index than SiN ($n$ = 2.0 at 1550 nm). From numerical simulations with various GaS positions on the SiN waveguide, the optical confinement factor was found to be relatively insensitive to the nanobelt position (see Supporting Information, Figure S2).

To demonstrate CW nonlinear optical operation, two CW lasers were used to excite SHG and sum-frequency generation (SFG). One pump wavelength ($\omega_1$) was fixed at 1550 nm, while the other ($\omega_2$) was tuned from 1560 nm to 1615 nm. Both pumps were coupled into the SiN waveguide in the in-plane direction, and the generated SHG and SFG signals were collected through an objective lens positioned above the GaS nanobelt-integrated waveguide (see experimental section and **Figure** S3 for details). Figure 5(f) shows the measured emission spectrum, which exhibited three distinct peaks corresponding to $SHG_1$ ($2\omega_1$), SFG ($\omega_1 + \omega_2$), and $SHG_2$ ($2\omega_2$) from shorter to longer wavelengths. As $\omega_2$ was varied, the peak positions of SFG and $SHG_2$ shifted accordingly, while $SHG_1$ remained fixed. The pump power dependence was further investigated, as shown in Figure 5(g), by fixing the power of $\omega_1$ and varying that of $\omega_2$. Under these conditions, the SFG intensity exhibited an approximately linear dependence on the $\omega_2$ pump power ($I_{SFG} \propto P^{\alpha}$ with $\alpha = 1.17 \pm 0.01$), whereas $SHG_2$ showed a near-quadratic dependence ($I_{SHG2} \propto P^{\alpha}$ with $\alpha = 2.46 \pm 0.06$), in good agreement with the theoretical expectations for SFG and SHG processes. These results clearly demonstrate wavelength conversion in a GaS nanobelt-integrated waveguide under CW laser excitation.

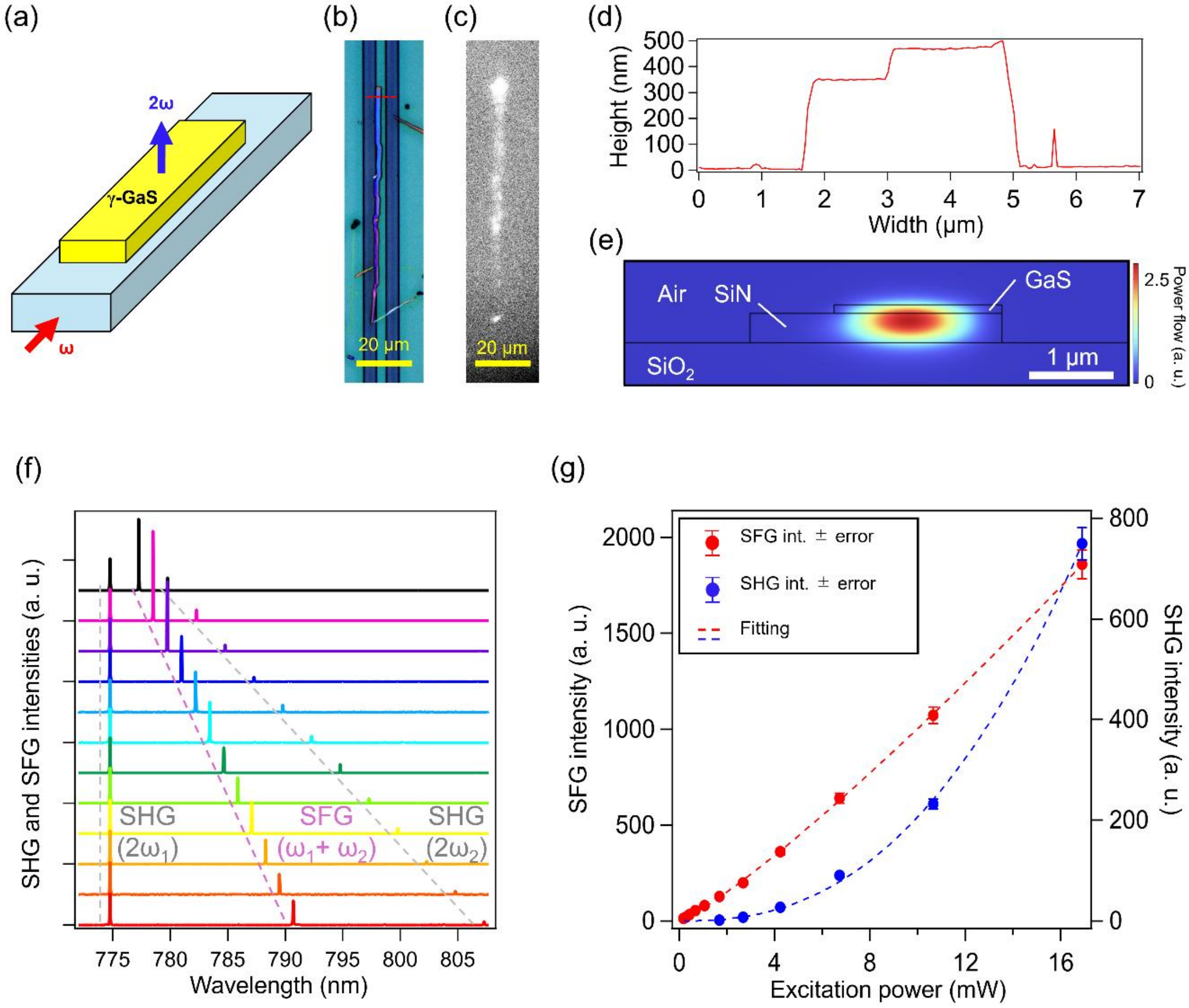

**Figure 5.** On-chip nonlinear wavelength conversion using GaS nanobelt-integrated SiN waveguide. (a) Schematic illustration of GaS nanobelt integrated onto SiN waveguide. (b) Optical microscope image and (c) SHG emission image of integrated device. SHG emission was obtained only under GaS nanobelt integrated area. (d) AFM line profile taken along dashed line in (b). (e) Simulated power flow distribution in waveguide. (f) SFG and SHG emission spectra under dual CW laser excitation. One pump wavelength ($\omega_1$) was fixed, while other ($\omega_2$) was tuned continuously. Dashed lines are eye-guide lines for each peak. (g) Pump power dependence of SFG and SHG intensities, confirming approximately linear and quadratic dependences, respectively.

## 3. Conclusion

We demonstrated that GaS nanobelts synthesized via self-catalyzed VLS growth predominantly crystallize in the non-centrosymmetric γ phase, overcoming the intrinsic centrosymmetry of bulk GaS. Atomic-scale structural analysis revealed that the in-plane orientation of the GaS unit layers remains highly oriented to a single direction, which is enabled by edge-selective growth kinetics at the Ga catalyst interface. This leads to deterministic in-plane dipole moment alignment across the entire nanobelt. As a result, the GaS nanobelts exhibit strong, reproducible second-harmonic generation with intensities comparable to those of GaSe. Furthermore, integration of the GaS nanobelts onto SiN waveguides enables efficient on-chip SHG and SFG under CW excitation. These findings establish γ-phase GaS nanobelts as a versatile and reliable one-dimensional nonlinear material for integrated photonic applications.

## 4. Experimental Section

**GaS Nanobelt Growth:**

GaS nanobelts were synthesized via self-catalyzed vapor–liquid–solid (VLS) growth using metalorganic chemical vapor deposition (MOCVD), following a procedure similar to that reported previously in our study.[32] Prior to the growth, sapphire (0001) substrates were annealed in an oxygen atmosphere at 1050°C for 6 h to obtain atomically flat surface terraces. Triethylgallium (TEGa) and diethylsulfide (DES) were used as the Ga and S precursors, respectively. A mixture of Ar and $H_2$ was used as the carrier gas. The growth temperature was maintained at 725°C.

**Fabrication of GaS Nanobelt-Integrated SiN Waveguides:**

SiN waveguides were fabricated as follows. A 350-nm-thick SiN layer was deposited on a $SiO_2$ (3 μm)/Si substrate by plasma-enhanced chemical vapor deposition. A resist layer was then spin-coated, and the waveguide pattern was defined by electron-beam lithography. Subsequently, the pattern was transferred into the SiN layer using $CHF_3$-based dry etching. Finally, the resist was removed to obtain the SiN waveguides.

GaS nanobelts were transferred onto the pre-fabricated SiN waveguides using a polymer-assisted transfer technique.[34] A polystyrene (PS) film was spin-coated onto the sapphire substrate on which the GaS nanobelts had been grown. Subsequently, KOH etching was performed to partially dissolve the sapphire surface and detach the GaS nanobelts together with the PS film. The resulting GaS nanobelt/PS film was then placed onto a glass slide. The glass slide was mounted on a micromanipulator, which was used to precisely control the position and in-plane orientation of the nanobelts during integration onto the SiN waveguide structures. After positioning the GaS nanobelt/PS film onto the SiN waveguides, the sample was heated to 140°C to melt the PS film and promote adhesion. Finally, the sample was immersed in toluene to dissolve the PS, yielding GaS nanobelt-integrated SiN waveguides.

**SHG Measurements for GaS Nanobelts, Exfoliated GaS and GaSe Flakes**:

As illustrated in Figure S3, SHG measurements were performed in a normal-incidence reflection geometry. For excitation, a Ti:sapphire laser with a wavelength of 810 nm was used as the fundamental light source, operating at a repetition rate of 80 MHz. The wavelength was converted to ~1500 nm via an optical parametric oscillator (OPO). The polarization of the excitation light was controlled using a half-wave plate, and unwanted polarization components were removed from the optical path by a polarizing beam splitter (PBS). Through an objective lens (OL), excitation light was focused on the samples. The generated SHG signal was collected by the same objective lens, and a dichroic mirror (DCM) reflected the SHG light into a spectrometer for spectral analysis.

**SHG and SFG Measurements for GaS Nanobelt-Integrated SiN Waveguide**:

As illustrated in **Figure** S4, a CW laser operating at a wavelength of 1550 nm was used as one excitation source. A second tunable CW laser was directed into an erbium-doped fiber amplifier (EDFA), and its power was adjusted using an attenuator (ATT), generating tunable radiation in the wavelength range of 1560–1615 nm. The two excitation beams were spatially combined using a beam splitter (BS) and then focused onto the GaS NB-WG. Nonlinear optical signals, including SHG and SFG generated from the GaS nanobelt, were collected along the surface normal using an OL. The emitted signals were subsequently directed into a spectrometer for spectral analysis.

**Data Availability Statement**

The data that support the findings of this study are available from the corresponding author upon reasonable request.

# Supplemental Information
# Non-Centrosymmetric γ-Phase GaS Nanobelts for On-Chip Nonlinear Photonic Applications

*Yukihiro Endo*[1]*, *Masato Takiguchi*[1, 2], *Yoshiaki Sekine*[1], *Hisashi Sumikura*[1, 2], *and Yoshitaka Taniyasu*[1]

[1]Basic Research Laboratories, NTT, Inc., 3-1 Morinosato Wakamiya, Atsugi, Kanagawa 243–0198, Japan
[2]NTT Nanophotonics Center, NTT, Inc., 3-1 Morinosato Wakamiya, Atsugi, Kanagawa 243–0198, Japan

**Contents**

## 1. SHG Spectra Measured from Large Number of GaS Nanobelts

The second-harmonic generation (SHG) responses of a sufficiently large number of GaS nanobelts were measured to evaluate sample-to-sample variations. As shown in **Figure S1**(a–c), 28 GaS nanobelts were randomly selected, as indicated by the yellow arrows. For each GaS nanobelt, SHG spectra were measured, as shown in Figure S1(d). A pulsed laser with a wavelength of 1530 nm was used as the excitation light. All measurements exhibited clear SHG peaks at 765 nm, indicating that all measured GaS nanobelts possessed non-centrosymmetric structures, in contrast to conventional bulk GaS.

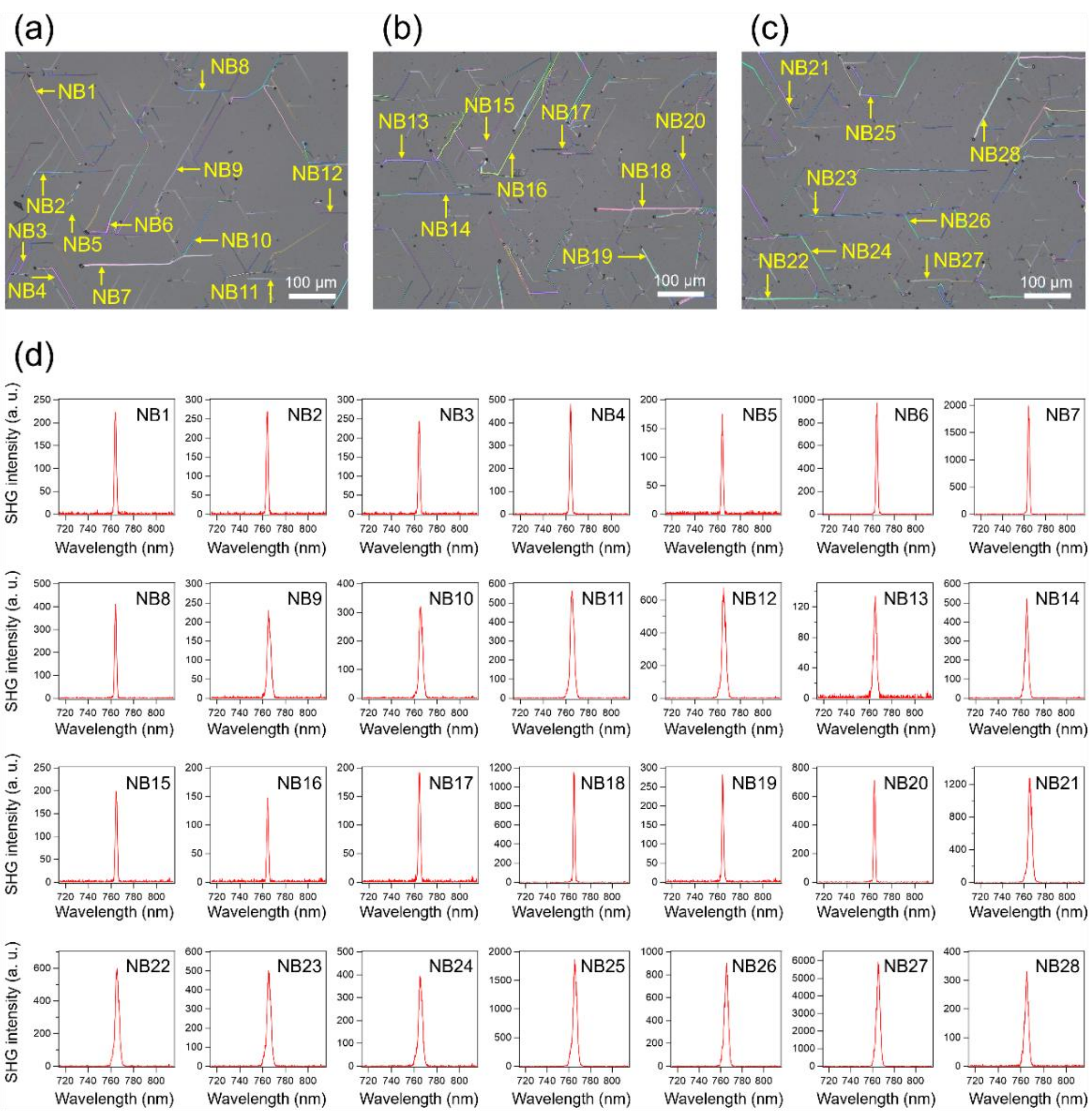

**Figure S1.** SHG spectra measured from large number of GaS nanobelts. (a–c) Optical microscope images of GaS nanobelts. 28 GaS nanobelts were randomly selected, as indicated by yellow arrows. (d) SHG spectra of GaS nanobelts. Each spectrum label corresponds to position marked in optical microscope images in panels (a–c).

## 2. Optical Simulations for GaS Nanobelt-Integrated SiN Waveguide

Numerical simulations were performed using COMSOL to evaluate the optical confinement in a GaS nanobelt-integrated SiN waveguide. The thicknesses of the GaS nanobelt and the SiN waveguide were set to 100 nm and 350 nm, respectively, while their widths were 2 μm and 3 μm. In the experiments, the position of the GaS nanobelt could deviate from the center of the SiN waveguide due to misalignment during the transfer process. To evaluate this effect, we carried out numerical simulations of the power flow in the waveguide cross-section while systematically varying the GaS position. The offset distance was defined as the displacement between the center of the SiN waveguide and that of the GaS nanobelt. The computational domain was set to 3 μm × 6 μm, and scattering boundary conditions were applied at the outer boundaries. The refractive indices were assumed to be $n_o \approx 2.64$ and $n_e \approx 2.35$ for GaS at 1550 nm, n = 2.0 for SiN, and n = 1.44 for $SiO_2$. An adaptive mesh was used, with a minimum mesh size of 31 nm in the GaS region. Here, the optical confinement factor in GaS was defined as $\int_{\mathrm{nb}} \varepsilon(r)|E|^2\, dS \;/\; \int_{\mathrm{all}} \varepsilon(r)|E|^2\, dS$. As shown in **Figure S2**(a–c), the propagating light follows the position of the GaS nanobelt. From this analysis, the optical confinement factor in the GaS nanobelt was found to be 11% and was insensitive to the nanobelt position [Figure S2(d)]. Thus, misalignment during the transfer process did not significantly affect the optical coupling.

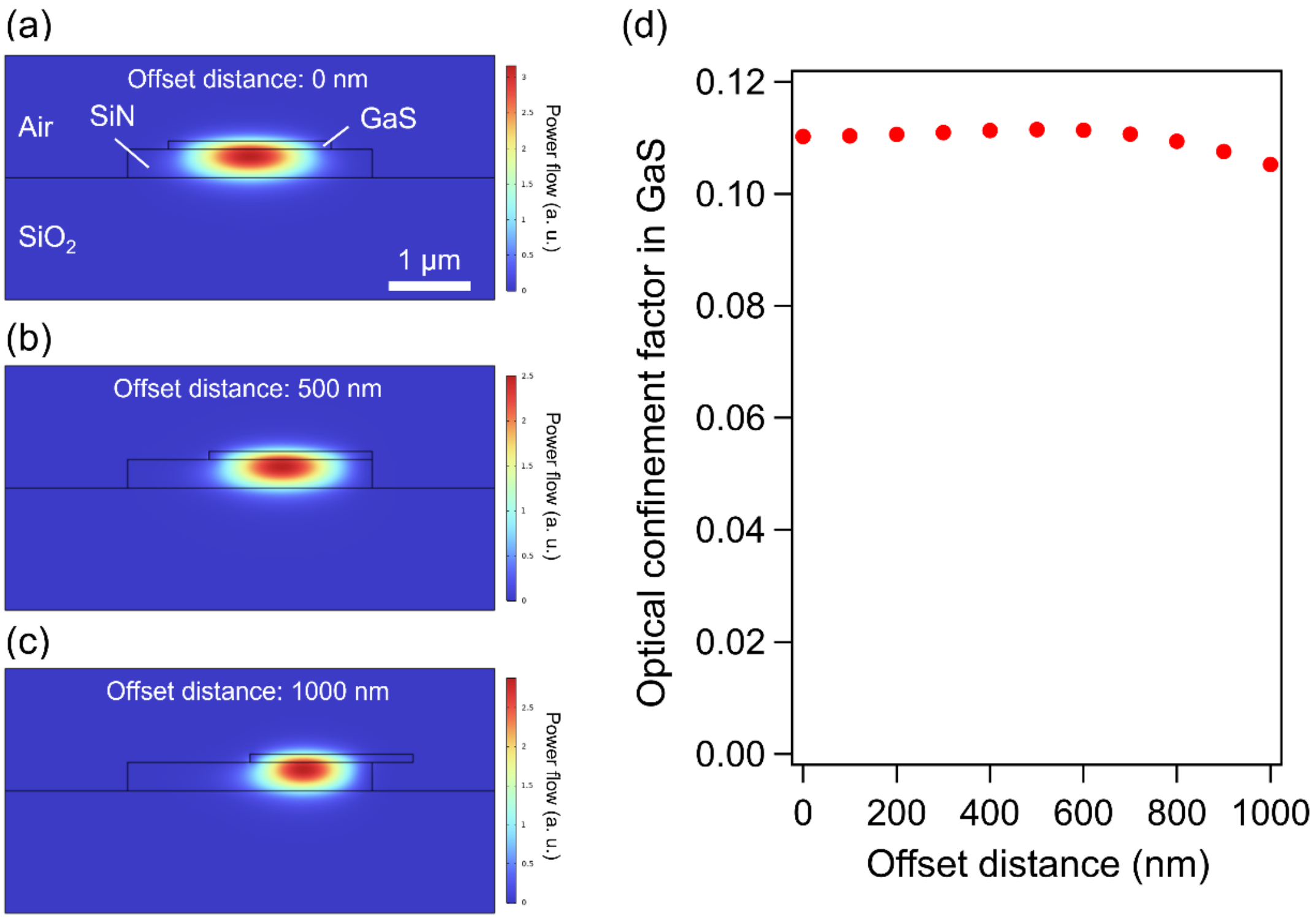

**Figure S2.** Simulated power flow in GaS nanobelt-integrated SiN waveguide for different nanobelt positions. (a–c) Cross-sectional power flow distributions, showing that optical modes follow nanobelt positions. (d) Optical confinement factors in GaS nanobelt as function of offset distance.

## 3. SHG Experiment Setup Using Pulsed Laser for Excitation

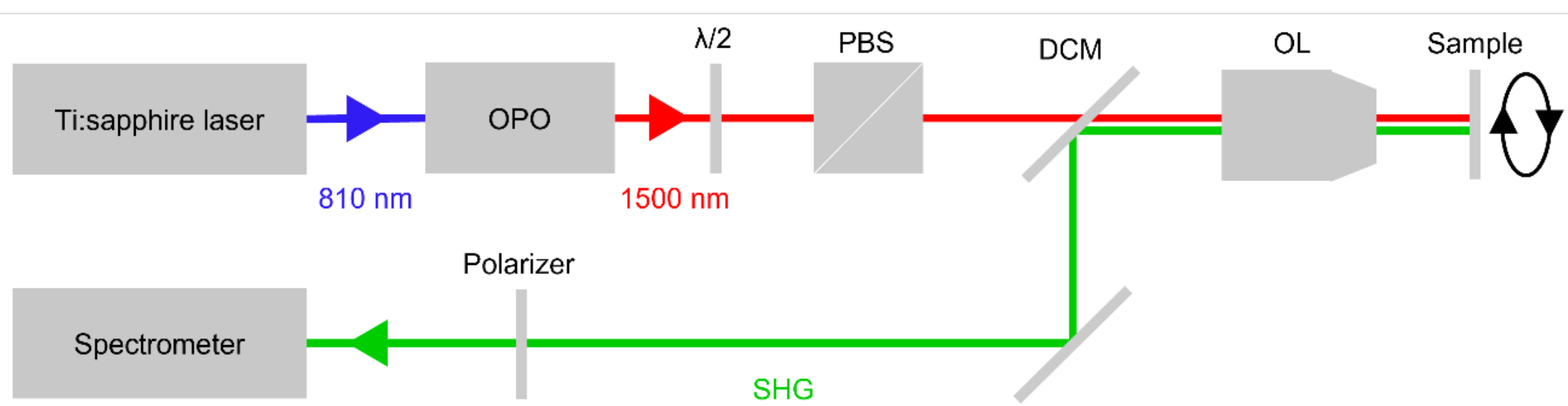

**Figure S3.** Schematic illustration of optical setup used for SHG measurements under normal-reflection geometry. Light paths for 810 nm, 1500 nm, and SHG signals are indicated by blue, red, and green lines, respectively.

## 4. SHG and SFG Experiment Setup Using Continuous-Wave Laser for Excitation

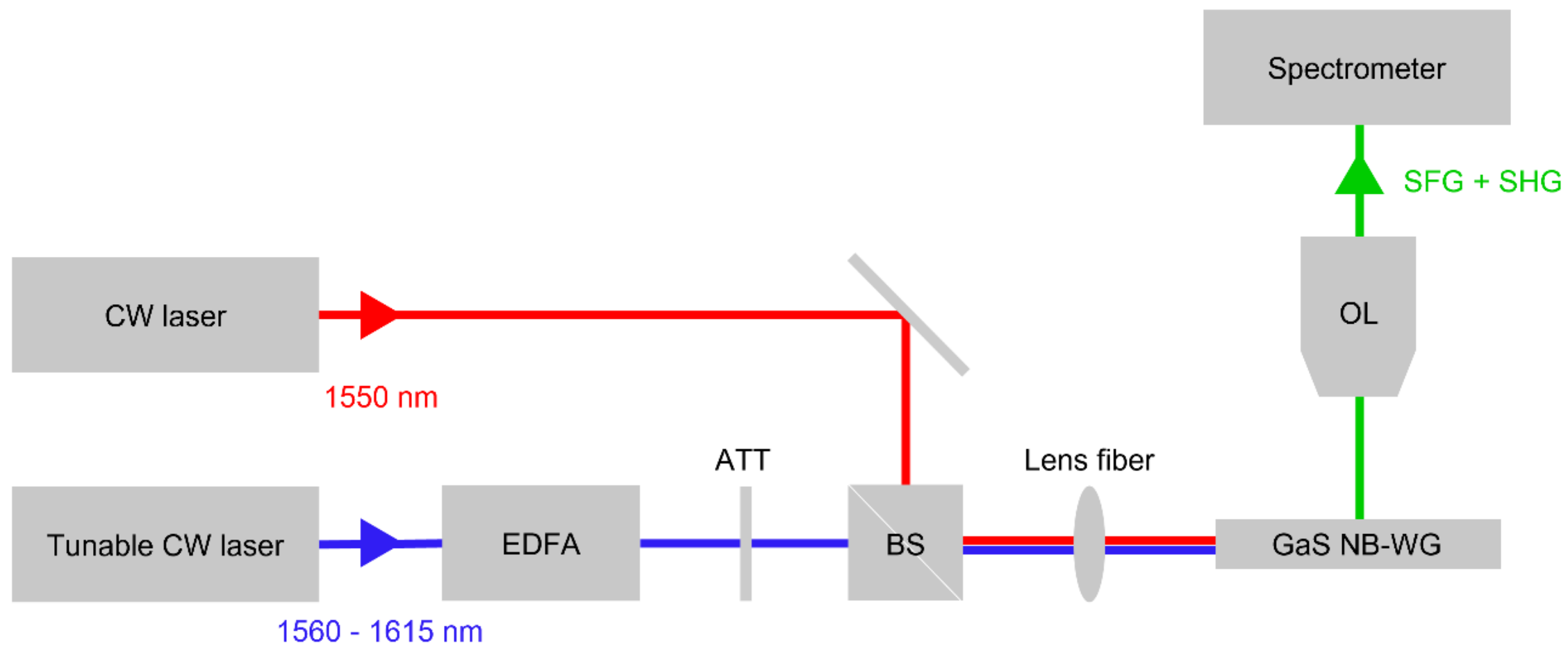

**Figure S4.** Schematic illustration of optical setup used for SHG and SFG measurements of GaS NB-WG. Light paths for 1550 nm, 1560–1615 nm, and SHG and SFG signals are indicated by red, blue, and green lines, respectively.